\documentclass{ws-procs9x6}

\def \gtsim    {\relax\ifmmode{\mathrel{\mathpalette\oversim >}}
                  \else{$\mathrel{\mathpalette\oversim >}$}\fi}
\def \ltsim    {\relax\ifmmode{\mathrel{\mathpalette\oversim <}}
                  \else{$\mathrel{\mathpalette\oversim <}$}\fi}
\def\oversim#1#2{\lower4pt\vbox{\baselineskip0pt \lineskip1.5pt
            \ialign{$\mathsurround=0pt#1\hfil##\hfil$\crcr#2\crcr\sim\crcr}}}
\makeatletter
\def\alt{\mathrel{\mathpalette\gl@align<}}
\def\agt{\mathrel{\mathpalette\gl@align>}}
\def\gl@align#1#2{\lower.6ex\vbox{\baselineskip\z@skip\lineskip\z@
\ialign{$\m@th#1\hfil##\hfil$\crcr#2\crcr\sim\crcr}}}
\begin{document}

\title{Some Results in M-Theory Inspired Phenomenology
\footnote{\uppercase{T}his work is supported  in part by a National Science Foundation Grant
PHY-0101015 and in part by  Natural Sciences and Engineering Research Council 
of Canada.}}

\author{R. Arnowitt,{$^\dagger$}
Bhaskar Dutta{$^*$},
and B. Hu{$^\dagger$}}

\address{ $^\dagger$Department of Physics, Texas A\& M University, \\
College Station, TX 77807, USA\\
$^*$Department of Physics, University of Regina, \\
Regina, Saskatchewan S4S 0A2, Canada}

\maketitle

\abstracts{{\it{It is a great pleasure to present this work in
honor  of Stan Deser whose numerous
fundamental contributions to general relativity and supergravity theory have
been so important in the development of high energy physics.}}\\\\
We consider string phenomenological models based on 11D Horava-Witten
M-theory with 5 branes in the bulk. If the 5-branes cluster close to the
distant orbifold plane ($d_n\equiv1-z_n\simeq 0.1$) and if the topological
charges of the physical plane vanish ($\beta^{(0)}_i=0$), then the Witten
$\epsilon$ terms (to first order) are correctly small and a qualitative
picture of the quark and lepton mass hierarchy arises without significant fine tuning.
If right handed neutrinos exist, a possible gravitationally induced cubic
holomorphic contribution to the Kahler potential can exist scaled by the
11D Planck mass. These terms give rise to Dirac neutrino masses at the
electroweak scale. This mechanism (different from the see-saw mechanism) is
seen to account for both the atmospheric and solar neutrino oscillations.
The model also gives rise to possible non-universal soft breaking $A$
parameters in the $u$ and $d$ second and third generation quark sector
($A_{2,3}^{(u,d)}$) which naturally can account for the possible
(2.4$\sigma$) break down of the Standard Model predictions in the recent B-factory
data for the
$B\rightarrow\phi K_s$ decays.
}

\section{Introduction}
With the advent of the ``landscape" in M-theory\cite{doug} with $10^{100}$(or
more!) possible string vacua, it is perhaps more important to try to use
phenomenology to help the development of string theory. Thus, for example, the
use of the ``experimental data" that life exists, i.e. the anthropic principle,
has begun to enter cosmology more seriously\cite{wein}. In particle theory string
phenomenology, one would like to construct a string model that at least semi
quantitativelygives rise to Standard Model(SM) physics we know to be true at low
energies. This means one wants to do more than just construct a theory that has
quarks and leptons arranged in three generations, but one would also like to
``explain" at least some of the things the SM can't explain, such as
\begin{itemize}
\item The quark and lepton mass hierarchy e.g. $m_u/m_t\simeq 10^{-5}$(where
$m_{(u,t)}$
are the (u,t) quark masses.
\item Supersymmetry(SUSY) soft breaking terms - which one are universal and
where non-universalities might occur.
\item The origin of neutrino masses
\end{itemize}
There are of course a huge number of string phenomenology models. We will try to
discuss these things here within Horava-Witten M-theory\cite{hw1,hw2,w1,h1} which
offers a framework which can allow some of the pecularities of the SM to emerge
naturally.
\begin{figure}
\centerline{\epsfxsize=6.1in\epsfbox{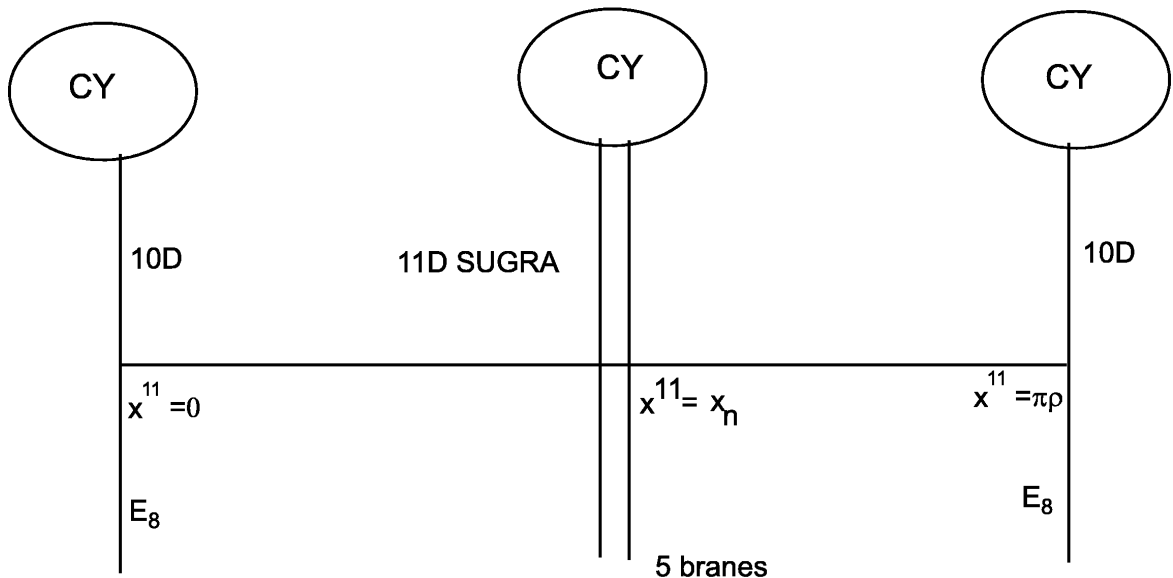}}   
\caption{Schemetic picture of H-W M-theory with 5 branes in the
bulk. 11D supergravity is in the bulk and $E_8$ Yang-Mills fields are on the 10D
orbifold planes.\label{fig1}}
\end{figure}

\section{Horava-Witten M-theory}
We summarize here some of the basic ideas of Horava-Witten (HW) M-theory. In HW
theory one considers an 11 dimensional(11D) orbifold, which to lowest order has
symmetry $M_4\times X\times S_1/Z_2$ where $M_4$ is Minkowski space and $X$ is a
Calabi-Yau(CY) 3-fold. In general, supersymmetry allows there to be a set of
5-branes orthogonal to the 11th coordinate x$^{11}$ wrapped  around the CY space with a
holomorphic curve.. The orbifold planes are at $x^{11}=0$ (the physical 10D plane) and $x^{11}=\pi\rho$
(the ``hidden sector"), shown schematically in Fig.1. To cancel 11D supergravity (SUGRA) anomalies,
there must be $E_8$ Yang-Mills gauge fields on the orbifold planes, and to make these supersymmetric
one must modify the Bianchi identity of the 5-form field strength G of 11D SUGRA to have sources on the
orbifold planes and 5-branes:
\begin{eqnarray}\label{eq1}
d G_{11 \, RSTU} &=& 4\sqrt{2}\pi(\frac{\kappa}{4\pi})^{2/3}
\left[J^{(0)} \delta (x^{11})\right.\\\nonumber &+&\left.J^{N+1}\delta
(x^{11}-\pi\rho)+\frac{1}{2}\Sigma_{n=1}^{N} J^{(n)} (\delta (x^{11}-x_n)+\delta
(x^{11}+x_n))\right]_{RSTU}.
\end{eqnarray}
where
\begin{eqnarray}\label{eq2}
J^{0} &=& -\frac{1}{16 \pi^2}\left( \mathrm{tr}\,F\wedge F -
\frac{1}{2}\,\mathrm{tr}\,R\wedge R \right)_{x^{11}=0}\\
J^{N+1} &=& -\frac{1}{16 \pi^2}\left( \mathrm{tr}\,F\wedge F -
\frac{1}{2}\,\mathrm{tr}\,R\wedge R \right)_{x^{11}= \pi\rho}
\end{eqnarray}
$J^{(n)}$, $n= 1\ldots N$ are  5-brane sources, $F^{(i)}$, $i=1,2$ are the $E_8$ Y-M field strengths,
$R$ is the curvature tensor and $\kappa^{-2/9}$ is the 11D Planck mass. The Bianchi identities then
imply.
\begin{equation}\label{eq3}
\Sigma^{N+1}_{n=0}\beta^{(n)}_i=0;\,\,\beta^{n)}_i\equiv\int_{C_i}J^{(n)},\,n=0,..,N+1.
\end{equation}
Finally the quantum theory is gauge invariant provided
\begin{equation}\label{eq4}
\lambda^2 = 2 \pi (4\pi\kappa)^{2/3}.
\end{equation}
where $\lambda$ is the 10D gauge coupling constant.

Thus HW M-theory represents a quantum theory based on the fundamental requirements of anomaly
cancelation, Yang-Mills gauge invariance and supergravity invariance.
While the $E_8$ Yang-Mills invariance guarantees grand unification of the gauge
couplings, the additional remarkable thing is that the quantum theory determines
the unified (10D) gauge coupling constant $\lambda$ in terms of the (11D)
graviational constant $\kappa$ as given in the Eq.(\ref{eq4}), and so gauge and
gravity are also unified.

Eq.(\ref{eq4}) leads to the Witten relations\cite{w1} for the GUT coupling
constant $\alpha_G$ and the Newton constant $G_N$
\begin{equation}\label{eq5}
\alpha_G=\frac{(4 \pi \kappa^2)^{2/3}}{2 \mathcal{V}};\,\,\,\,
G_N=\frac{\kappa^2}{16 \pi^2 \mathcal{V}\rho}
\end{equation}
where $\mathcal{V}\equiv (M_G)^{-6}$ is the CY volume. Assuming 
$M_G$ is the GUT mass, $M_G\simeq 3\times 10^{16}$ GeV (since experimentally
grand unification should occur at the compactification scale) one has ($\alpha_G
\simeq 1/25$)
\begin{equation}\label{eq6}\kappa^{- 2/9} \cong 2M_G,\,\, (\pi\rho)^{-1} \cong 4.7 \times
10^{15}{\rm GeV}\end{equation}
Eq.(\ref{eq6}) implies two points. First since the 11 dimensional gravitational
mass $\kappa^{- 2/9}$ is the fundamental mass of the theory, one sees that $M_G$
rather than the Planck mass from $G_N$ is the fundamental mass scale. ($M_{pl}$
is a derived 4D quantity from Eq.(\ref{eq5}) which is accidentally large.)
Second the orbifold length $\pi\rho$ is $O(10)$ times  larger than the CY size
$M_G^{-1}$. Thus Witten discusssed a solution in terms of an expansion in
powers of $\kappa^{2/3}$ or more explicitly in powers of the dimensionless
parameter
\begin{equation}\label{eq7}
\epsilon=(\frac{\kappa}{4\pi})^{2/3}\frac{2\pi^2\rho}{\mathcal{V}^{2/3}}
\end{equation}and this has been extensively examined to $O(\epsilon)$.
(See\cite{low4} and references therein). Actually $\epsilon$ is not small,
i.e. $\epsilon\simeq$ 0.9. However for the HW models we will consider $\epsilon$
is multiplied by a small parameter $d_n\simeq 0.1$ so that $\epsilon d_n$ is
indeed small.  

Chiral matter arises from expanding the Y-M field strength $F_{\mu{\bar b}}$
($\mu$=0, 1, 2, 3 is in Minkowski space and $b$, ${\bar b}$ in CY space) in CY (0,1)
harmonic functions $u_{I{\bar b}}$:
\begin{equation}\label{eq8}
F_{\mu \bar b}=\sqrt{2\pi\alpha_G}\Sigma_I u_{Ib}T_xD_{\mu}C^I(x)
\end{equation}
Here $C^I(x)$ are the chiral fields, $I$ is the family index, $T_x$ is a group
generator. (Thus gauge and chiral matter are also unified.) The quantities
needed to construct a phenomenological model to $O(\epsilon)$ (following the
analysis of \cite{low4}) are the gauge function on the physical orbifold plane
\begin{equation}\label{eq9}
f^{(1)}=S+\epsilon T^i \left
(\beta^{(0)}_i+\sum^{N}_{n=1}(1-z_n)^2\beta^{(n)}_i \right ),
\end{equation}
the matter Kahler metric ($K=Z_{IJ}\bar {C^I}C^J$)
\begin{equation}\label{eq10}
Z_{IJ}=e^{-K_T/3}\left [G_{IJ}-\frac{\epsilon}{2V}{\Gamma}^i_{IJ}
(\sum^{N}_{n=1}(1-z_n)^2\beta^{(n)}_i +\beta^{(0)}_i)\right ]
\end{equation}
and the Yukawa coupling constants
\begin{equation}\label{eq11}
Y_{IJK}=2\sqrt{2\pi\alpha_G}\lambda_{IJK}\int_X\Omega\wedge u_{I}\wedge  u_{J}\wedge
u_{K}
\end{equation}
Here $V=Re S$ is the CY volume modulus, $R b_i\equiv Re T_i$ (R is the orbifold
modulus),
\begin{equation} \label{eq12}
G_{IJ}(a^i;R)=\frac{1}{{\mathcal{V}}}\int_X\sqrt{g}g^{a\bar
b}u_{Ia}u_{J\bar b}
\end{equation}
$\Omega$ is the covariantly constant (0,3) form, $z_n=x_n/{\pi\rho}$ are the
positions of the 5-branes and $X$ is the CY space. (Explicit forms for $K_T$ and
$\Gamma^i_{IJ}$ can be\cite{h1} found in \cite{low4}.) Thus $f^{(1)}$ and $Z_{IJ}$ contain a
zeroth order piece and an $\epsilon$ correction. Supersymmetry is broken by the
Horava mechanism \cite{h1} of a condensate on the distant orbifold plane with
superpotential.
\begin{equation}\label{eq13}
W=h exp[-\frac{6\pi}{b\alpha_G}(S+\epsilon T^i [-\beta^{(0)}_i+\sum^{N}_{n=1}(z_n^2-1)\beta^{(n)}_i ]),
\end{equation}
where $h\simeq \alpha_G/\mathcal{V}$ and $b$=90 (for the $E_8$ gauge group on
the distant brane).
\section{Quark and Lepton Masses}
In conventional SUGRA GUT models, one assumes a complicated Yukawa matrix at
$M_G$ and calculates the masses and CKM matrix at the electroweak scale. Thus for
example, a u-quark Yukawa matrix at $M_G$ which accounts approximately for the
low energy observed electroweak data is given in Fig.2\cite{rrr}. 
\vspace{1cm}\hrule{}\begin{displaymath}
Y_U=\left(\begin{array}{ccc} 0 & \sqrt{2}\lambda^{6}  & 0 \cr
\sqrt{2}\lambda^{6} & \sqrt{3}\lambda^{4} & \lambda^{2} \cr 0  &
\lambda^{2}& 1 \end{array}\right).
\end{displaymath}
Fig.2. The u-quark Yuakawa matrix at $M_G$, where $\lambda=0.2$ is the
Wolfenstein parameter.\hrule{}\vspace{1cm}

However,
$\lambda^6\simeq 10^{-5}$ is needed to obtain $m_u/m_t\simeq 10^{-5}$ at the
electroweak scale, so that the quark mass hierarcy is not explained. In HW theory,
the Yukawa matrices are integrals over the CY space, Eq.(\ref{eq11}), with no
reason to assume that some components should be anomalously small. However, this
is not necessarily the case for the Kahler metric for Eq.(\ref{eq10}). $Z_{IJ}$
is given there in an expansion in $\epsilon$. As mentioned above, $\epsilon$ is
not too small ($\epsilon\simeq 0.9$) making the validity of the expansion in
doubt. However, if it were possible to have a CY manifold where
$\beta^{(0)}_i$=0, then what actually appears in Eqs.(\ref{eq9}) and (\ref{eq10})
are the combinations
\begin{equation}\label{eq14}
\epsilon d_n^2,\,\,\,\epsilon d_n;\,\,d_n\equiv 1-z_n
\end{equation}
which would be small if $d_n\simeq 0.1$. Thus the $\epsilon$ expension would be
reasonable (at least in first order) if the 5-branes in Fig.1 all cluster near
the distant orbifold plane $x^{11}=\pi\rho$ and the topological parameters
$\beta^{(0)}_i$
 vanish. We will see below that this structure needed for convergence of the $\epsilon$ expansion naturally gives 
 rise to the quark
 and lepton mass hierarchy.
 
 However, the condition $\beta^{(0)}_i$=0 is highly non trivial, and it is
 possible to show this is impossible for an elliptically fibered CY
 manifold\cite{lo}. This no-go theorem can be evaded, however  using a torus fibered CY
 (with two sections) and one can show that a three generation model with
 $\beta^{(0)}_i$=0 and SU(5) symmetry satisfying all topological constraints
 exists provided one uses a del Pezzo base $dP_7$\cite{ab}. (More recently a
 similar result with SO(10) symmetry has been obtained \cite{Far}) The torus
fibration also allows for a Wilson line breaking of the $SU(5)$[or $SO(10)$]
symmetry at $M_G$ to the Standard Model, giving rise to the phenomenologically
desired grand unified supergravity GUT model below $M_G$.

In the following, we will treat $q\equiv(u_L,\,u_R,\,d_L)$ together [as in
SU(5)], and $d_R$ seperately. The non-zero elements of $G_{IJ}$ are a priori
expected to be $O(1)$.  A quark mass spectrum which is qualitatively right can
emerge if we assume that the $G_{IJ}$ contributions to $Z_{IJ}$ contribute to
all generations for the $d_R$ states but only to the first two generations for
the $q$ states. The general form then for $Z^q$ is shown in Fig.3. The contributions
involving the third generations of $Z^q$ must be small since $d_n$ must be small for the Witten expansion to converge.
\vspace{1cm}\hrule{}\begin{displaymath}
Z^q=\left(\begin{array}{ccc} O(1) & O(1)  & O(d_n^2) \cr
O(1) & O(1) & O(d_n^2) \cr O(d_n^2)  &
O(d_n^2)& O(d_n^2) \end{array}\right).
\end{displaymath}
Fig.3. The qualitative form of the Kahler matrix $Z^q$ for
$q\equiv(u_L,\,u_R,\,d_L)$ where $d_n\equiv 1- z_n\simeq 0.1$\hrule{}\vspace{1cm}

 The superpotential for the chiral matter fields has the form 
\begin{equation}\label{eq15}
W_Y=e^{\frac{1}{2}K}\frac{1}{3}Y_{IJK}C^IC^JC^K
\end{equation} where $Y_{IJK}$ is given in Eq.(12). For example the $q_Lu_RH_2$
coupling [$q_L=(u_L,d_L)$] (which at the electroweak scale gives rise to the
u-quark mass) is
\begin{equation}\label{eq16}
W_Y=\frac{1}{4R^{3/2}V^{1/2}}(Y^{(u)} q_L H_2 u_R); \, V\equiv r^6.
\end{equation} where we parametrize $V$ by $r^6$. To obtain the canonical form,
one must reduce $Z_{IJ}$ to a unit matrix. Thus the canonical matter variables
are \begin{equation} C^{I'}={\frac{1}{\sqrt{f_T}}}(U S)_{IJ}
{C^{J}}^{\prime};\,\,f_T=e^{-K_T/3} 
\label{eq17}\end{equation} where $U$ diagonalizes $Z_{IJ}$ and $S$ reduces it to
a unit matrix
\begin{equation} {\rm diag} S= (\lambda_1^{-1/2},\, \lambda_2^{-1/2},\,\lambda_3^{-1/2}).
\label{eq18}\end{equation} Here $\lambda_i^{(u,d)}$ are the eigenvalues of
$Z_{IJ}$. In terms of canonical variables then, $W_Y=u_L^{\prime} \lambda^{(u)}
u_R^{\prime} H_2^{\prime}$ where
\begin{equation}\label{eq19}
    \lambda^{(u)}_{IJ} = \frac{1}{8 \sqrt{2}}\frac{1}{R^3V^{1/2}}\frac{1}{\sqrt{G_{H_2}}}
                      (S^{u}{\tilde U^{u}}Y^{u}U^{u}S^{u})_{IJ}
\end{equation}
where $H_2^{\prime}=\sqrt{G_{H_2}} H_2$. Since unitary matrices have entries of
$O(1)$, the scale of the u-quark masses are qualitatively governed by the
eigenvalues of the Kahler matrix i.e.
\begin{equation}\label{eq20}
    m_u:m_c:m_t=\frac{1}{\lambda^u_1}:\frac{1}{\lambda^u_2}:\frac{1}{\lambda^u_3}
\end{equation}
As an example, consider $Z^u$ to be 
\begin{equation}Z^u=f_T\left(\begin{array}{ccc} 1  & 1/3  & 0 \cr 1/3 & 1/8 & (2/3) d_n^2 \cr
 0  &
(2/3) d_n^2& (1/3) d_n^2 \end{array}\right);\,\,d_n=1-z_n=0.1.
\label{eq21}\end{equation}\vspace{0.2cm}
Then \begin{equation}
 \lambda^u_1=1.12;\,\,\lambda^u_2=0.0144;\,\,\lambda^u_3=1.05 d_n^4
 \label{eq22}\end{equation} and so
\begin{equation}
 m_t\simeq100 m_c\simeq (100)^2m_u
 \label{eq23}\end{equation} which is qualitatively correct.

Thus the quark masses hierarchy arises from the effects of the HW 5 branes on
the Kahler metric when the 5-branes lie close to the distant orbifold plane. The
parameter of smallness in the quark mass matrix is then the geometrical one
$d_n\equiv 0.1$.
\begin{table}
\vspace{0.2cm} \hrule{}
\begin{eqnarray}\nonumber
Z^u=f_T\left(\begin{array}{ccc} 1  & 0.3452  & 0 \cr 0.3452 & 0.1311 &
0.006365 \cr 0  & 0.006365 & 0.00344 \end{array}\right); \,\,\,
Z^d=f_T\left(\begin{array}{ccc} 1 & 0.496  & 0 \cr 0.496 & 0.564 &
0.435 \cr 0  & 0.435 & 0.729  \end{array}\right);
\end{eqnarray}
\begin{eqnarray}\nonumber
Z^l=f_T\left(\begin{array}{ccc} 1  & -0.547  & 0 \cr -0.547 & 0.432 &
0.025 \cr 0  & 0.025 & 0.09 \end{array}\right); \,\,\,
Z^e=f_T\left(\begin{array}{ccc} 1 & 0.624  & 0 \cr 0.624 & 0.397 &
0.00574 \cr 0  & 0.00574 & 0.004407  \end{array}\right).
\end{eqnarray}
\vspace{0.2cm}\begin{flushleft}{Fig.4 Kahler matrices $Z^{(u,d,l,e)}_{IJ}$ and
Yukawa matrices  $Y^{(u,d,e)}$ for $\tan\beta$=40.}\end{flushleft}\hrule{}\end{table}
\begin{table}
\vspace{0.2cm} \hrule{}
\begin{center} \begin{tabular}{|c|c|c|}
 \hline Quantity&Theoretical Value&Experimental Value\\\hline
$m_t$(pole)& 175.2 & $174.3 \pm5.1$ \\
$m_c$($m_c$)& 1.27 & 1.0-1.4\cr
$m_u$(1 GeV)& 0.00326 & 0.002-0.006\\
$m_b$($m_b$)& 4.21 & 4.0-4.5\cr
$m_s$(1 GeV)& 0.086 & 0.108-0.209\\
$m_d$(1 GeV)& 0.00627 & 0.006-0.012\\
$m_\tau$& 1.78 & 1.777\\
$m_\mu$& 0.1054 & 0.1056\\
$m_e$& 0.000512 & 0.000511\\
$|V_{us}|$& 0.221 & $0.2210\pm 0.0023$ \\
$|V_{cb}|$& 0.042 & 0.0415$\pm$0.0011 \\
$|V_{ub}|$& $4.96 \times 10^{-3}$ & $3.80^{+0.24}_{-0.13} \pm 0.45 \times 10^{-3}$ \\
$|V_{td}|$& $6 \times 10^{-3}$ & $9.2 \pm 1.4 \pm 0.5 \times 10^{-3}$ \\
$\sin 2\beta$ & 0.803 &  $0.731\pm 0.056$ \cite{hfag}\\
\hline\end{tabular}\end{center} \begin{flushleft}{ Fig. 5. Quarks and leptons masses and CKM matrix
elements obtained from the model of Fig.4\cite{abh}. Masses are in GeV.
Experimental values for lepton and quark masses are from \cite{PDG02a}
and CKM entries from \cite{lp03} unless otherwise noted. The only mass
input in the second column is $m_t$(pole) which sets the scale of all
other masses.}\end{flushleft}\hrule{}\vspace{1cm}
\end{table}
A similar analysis holds for d-quarks and charged leptons, and a
semi-quantitative picture of their mass ratios can also be achieved naturally
using only $d_n=0.1$. Of course, to precisely get the quark and lepton masses,
one must accurately chose $Z^{u,d,e}$, $Y^{u,d,e}$ and  one can check that this
is possible. Thus Fig.4, in accord with the general HW assumptions stated above,
gives rise to the table in Fig.5 for the quark, lepton and CKM elements\cite{abh}. Fig.4
determines all the mass ratios and CKM elements and the absolute values of all
quark and lepton masses are then obtained by putting in only one mass, i.e.
$m_t(pole)=175.2$ GeV. The actual value of $m_t$ can be related to the CY moduli
parameters $R$ and $r$. Thus using Eqs.(17) and (20) for the t-quark gives the
relation (for $m_t=175$ GeV),
\begin{equation}
Rr(G_{H_2})^{1/2}=6.82;\,\,m_t=175 {\rm GeV}
 \label{eq24}\end{equation}
which is a reasonable condition one might expect to hold for the moduli of the
physical CY manifold.
\section{Neutrino Masses}
The low energy Standard Model does not allow for neutrino masses. The conventional way of
introducing neutrino masses is to assume that they arise from GUT scale physics
by the see-saw mechanism. Here one assumes the existence of right-handed neutrinos $\nu_R$
which develop a GUT size
Majorana mass (M) as well as an electroweak scale size Dirac mass (m) with the
physical left-handed neutrinos $\nu_L$. Eliminating the heavy $\nu_R$ fields,
one is left with Majorana masses for $\nu_L$ of size $m^2/M$, which is the size
observed experimentally.

The Horava-Witten model offers an alternate possibility for neutrino masses
arising from the Kahler potential\cite{abh}. Along with the quadratic matter Kahler
potential discussed in Sec.3, the Kahler potential can in principle have
gravitationally induced cubic terms which would be scaled on dimensional grounds 
by $1/\kappa_{11}\simeq
M_G$ (the 11D Planck mass). In general these terms are negligible. An exception arises if
there are holomorphic terms involving the $\nu_R$ neutrinos. The only gauge invariant holomorphic cubic
terms involving $\nu_R$ is
\begin{equation}\label{eq25}
K_{\nu}=\kappa_{11} Y^{(\nu)} l_L H_2 \nu_R
\end{equation}
where $l_L=(\nu_L,\, e_L)$ and $Y^{(\nu)}$ is a neutrino Yukawa matrix. (The
possibility of forming this structure is the unique feature of the SUSY SM that
$l_L$ and $H_2$ poseess the same $SU(2)_L$ quantum numbers.) 
By a Kahler transformation one can move this term into the superpotential,
\begin{eqnarray} \label{eq26}
K & \rightarrow  & K - K_{\nu},  \nonumber \\
W & \rightarrow  & e^{\kappa_4^{\;2} K_{\nu}} W = W +
\kappa_4^{\;2} K_{\nu} W + \cdots
\end{eqnarray}
where $1/\kappa_4$ is the 4D Planck mass. When supersymmetry breaks, one has the
additional superpotential term at $M_G$ of 
\begin{equation}\label{eq27}
W^{\nu}= \left\langle W \right\rangle\frac{\kappa_4^{\;2}}{M_G} Y^{(\nu)} l_L H_2 \nu_R
\end{equation}
which leads to Dirac masses at the electroweak scale  when $H_2$ grows a VEV.

To find the size of the neutrino masses, one must proceed as in the squark and
lepton case transforming to canonical fields. $W^\nu$ becomes
\begin{equation}\label{eq28}
W^{\nu}=\nu_L^{\prime} \lambda^{(\nu)} \nu_R^{\prime} H_2^{\prime}
\end{equation}
where 
\begin{equation}\label{eq29}
    \lambda^{(\nu)}_{IJ} =
    \frac{1}{\sqrt{2}}\frac{1}{R^{3/2}}\frac{1}{\sqrt{G_{H_2}}}\frac{\kappa_4^2 \left\langle W \right\rangle}{M_G}
                     (S^{(l)}{ U^{(l)}}Y^{(\nu)}U^{(\nu)}S^{(\nu)})_{IJ}
\end{equation}
Here $S^{(l)} U^{(l)}$ diagonalizes the charge lepton Kahler matrix (and was determined in Sec.3
to obtain the correct lepton masses) while $U^{(\nu)}S^{(\nu)}$ diagonalize the $\nu_R$ Kahler
matrix $Z^{\nu_R}$. As an example we consider $\tan\beta=40$ with $Z^{\nu_R}$ and $Y^{\nu}$ of
Fig.6 \cite{abh} which has the 5-brane induced structure as in $Z^q$. 
\vspace{1cm}\hrule{}\begin{displaymath}
Z^{\nu_R}=\left(\begin{array}{ccc} 1  & -0.465  & 0 \cr -0.465 &
0.3105 & 2.54 d_n^2 \cr 0  & 2.54 d_n^2 & 2.7 d_n^2 \end{array}\right).
\end{displaymath}
\begin{displaymath}
diag Y_{\nu}=\left(4,\,0.4,\,4\right).
\end{displaymath}
Fig.6. $Z^{\nu_R}$ and $Y^{\nu}$ for the case $\tan\beta=40$. As in the quark lepton sector
$d_n=0.1$.\hrule{}\vspace{1cm} Using the renormalization group equations(RGE), one finds at the electroweak
scale that $U_{e3}$ is small
\begin{equation}\label{eq30}
|U_{e3}|=0.005
\end{equation} and hence the solar($S$) neutrino oscillations are governed by  $\Delta m_{21}^2$and the
atmospheric(A) oscillations by $\Delta m_{32}^2$. Thus for the solar oscillations 
we find\cite{abh}
\begin{equation}\label{eq31}
\Delta m_{21}^2= 5.5 \times 10^{-5} \,\, \rm{eV}^2; \,\,\tan^2 \theta_{12} = 0.42.
\end{equation}
in accord with the KAMLAND data\cite{solar}
\begin{equation}\label{eq32}
5.6\times 10^{-5}\le
\Delta m_S^2/\rm{eV}^2\le 8.9\times 10^{-5}
\end{equation}
and for the atmospheric oscillation\begin{equation}\label{eq33}
\Delta m_{32}^2= 2.7 \times 10^{-3} \,\, \rm{eV}^2; \,\,\tan^2 \theta_{23} = 0.93.
\end{equation}
in accord with the SuperKamiokande and K2K large mixing angle solution\cite{atom}
\begin{equation}\label{eq34}
0.85\le \sin^2 2\theta_A \le 1;\,\,\,\,\,1.4\times 10^{-3}\le\,\,
\Delta m_A^2/\rm{eV}^2\le 3.8\times 10^{-3}.
\end{equation}
The mass scale of the neutrino masses is determined by the front factor of Eq.(\ref{eq29}),

\begin{equation}\label{eq35}
    Q=\frac{\kappa_4^2 \left\langle W \right\rangle}{R^{3/2}\sqrt{G_{H_2}}M_G}
\end{equation}
and the 
above neutrino masses are obtained by choosing\begin{equation}\label{eq36}
    Q=1.072\times 10^{-14}
\end{equation}
However, there are theoretical constraints on $Q$, as $\left\langle W \right\rangle$ is related
to $m_{3/2}$, and one might ask whether Eq.(\ref{eq36}) is a reasonable value for Q. One has
that\begin{equation}\label{eq37}
   m_{3/2}=\frac{5}{\sqrt{12}}\frac{\kappa_4^2 \left\langle W \right\rangle}{R^{3/2}}
\end{equation}
Hence, \begin{equation}\label{eq38}
   Q=\frac{\sqrt{12}}{5}\frac{1}{\sqrt{G_{H_2}}}\frac{m_{3/2}}{M_G}
\end{equation}
For $m_{3/2}$=500 GeV, one finds that the choice Eq.(\ref{eq36}) (which gives the correct size
of neutrino masses) is satisfied by:
\begin{equation}\label{eq39}\sqrt{G_{H_2}}=1.077\end{equation}
a reasonable result. Thus the smallness of neutrino masses in the model, i.e. the smallness of
$Q$ in Eq.(\ref{eq36}), is related to the gauge hierarchy i.e. that $m_{3/2}/M_G\simeq 10^{-14}$ 
 in Eq.(\ref{eq38}). Using
the previous result from the t-quark mass, Eq.(\ref{eq24}), one then finds
\begin{equation}\label{eq40}
    R=\frac{6.74}{r}
\end{equation}
relating the orbifold and CY moduli.

However, one can go further, since $\left\langle W \right\rangle$ in Eq.(\ref{eq37}) can be
obtained from Eq.(14):
\begin{equation}\label{eq41}
   m_{3/2}=\frac{5}{\sqrt{12}}(\frac{r}{6.74})^{3/2}\kappa_4^2 \frac{\alpha_G}{\mathcal{V}}
   exp[-\frac{6\pi}{b\alpha_G}(r^6-2\epsilon T_i\Sigma d_n\beta_i^{(n)})]
\end{equation}
For $m_{3/2}$=500 GeV, Eq.(\ref{eq41}) determines $r$, and using Eq.(\ref{eq40}) one finds
\begin{equation}\label{eq42}
   r\simeq1.34;\,\,\,R\simeq 4.97\,\, {\rm for \,\,m_{3/2}=500\,GeV,\,\,\tan\beta=40}
\end{equation}
Thus the model we have constructed giving rise to the quark, lepton, neutrino mass hierarchies
implies that the Calabi-Yau and orbifold moduli posseses reasonable values.
\section{CP Violation in B Decays  and Non-Universal Soft Breaking}

In mSUGRA GUT models, one assumes that universal soft breaking of SUSY occurs at $M_G$ so
that all scalar particles have a common mass $m_0$ at $M_G$, and the cubic soft breaking
mass $A_0$ is common for all Yukawa couplings $Y_{IJK}$ at $M_G$. This in general is a good
fit to all data with one possible exception. Recent measurements at the B factories (BaBar
and Belle) for a class of B decays $B\rightarrow\phi K^0_s$, $B\rightarrow\phi K^0_L$ and 
$B\rightarrow\eta' K^0$ etc., have indicated a possible breakdown of the Standard Model.
These decays are unique in two ways. First the SM tree level contribution is zero and so
both the SM and SUSY contributions begin at the first loop level. Since both contributions
are a priori of comparable size, these decays are a place where the new physics of SUSY might
be experimentally observable. Second, at the quark level, all these decays have the general
structure of \begin{equation}\label{eq43}
   b\rightarrow s+\bar s +s
\end{equation}so we are dealing with a well defined class of processes.

B decays exhibit large CP violation and so the B factories can measure both the CP
violating phase $\beta$ as well as the branching ratios. In the SM, CP violation stems from
a single source, the phase in the CKM matrix. The CP violating phase $\beta$ in B decays is
governed by the CKM phase and in the SM one has $\sin2\beta\simeq0.73$. However, SUSY
allows for additional CP violating phases which could modify the value of $\sin2\beta$.
Averaged over all the $b\rightarrow s$ decays, BaBar finds a 2.7$\sigma$ deviation of
$\sin2\beta$ from the SM results\cite{babar} and Belle a 2.4 $\sigma$ deviation\cite{belle}. The decay mode
that is most theoretically accessible is $B\rightarrow\phi K^0_S$. Averaging the BaBar and
Belle results for this decay, we find $\sin2\beta_{\phi K_s}=0.18\pm 0.23$ which is
$2.4\sigma$ deviation from the SM. In addition, the measured branching ratio for both 
$B^0\rightarrow\phi K^0$ and $B^+\rightarrow\phi K^+$ are large. Thus for the well measured $B^+$
decay one has 
Br$[B^+\rightarrow\phi K^+]$=$(10.9\pm1.0)\times 10^{-6}$, while over most
of the parameter space the SM expects a branching ratio of $5\times 10^{-6}$, 
much lower than the experimental
result.

While the errors are still quite large (and B decay calculations are very complicated) one
might ask how one could account for such effects. It turns out that mSUGRA (with universal
soft breaking) gives result almost identical to SM[abh1], so if the above results are borne out
by further data, they would imply both a breakdown of the Standard Model and of mSUGRA! For
a SUGRA GUT  model, then the only way one might account for this data would be by assuming
non-universal soft breaking at $M_G$. If the soft breaking masses have natural (i.e. electroweak) size,
then the only non-universal terms that can account for the B decays data is a non-universal
cubic soft breaking $A$ parameter mixing the second and third generations in the $u$ or $d$
quark sectors. Detailed calculations\cite{abh1} then show that\begin{equation}\label{eq44}
  |A_{23}^{u,d}|\simeq A_0
\end{equation}
can correctly give both the experimental $\sin2\beta_{\phi K_s}$ and the  $B\rightarrow\phi
K$ branching ratios.

It is interesting that the Horava-Witten model we have been considering produces
non-universal $A$ parameters in precisely the $A_{23}^{u,d}$ sector. The effective
potential for the $A$ soft breaking term is $V_A=A_{IJK}C^IC^JC^K$ or 
\begin{equation}\label{eq45}
  V_A=A_{IJK}C^{I'}C^{J'}C^{K'}Q^I_{I'}Q^J_{J'}Q^K_{K'}
\end{equation}where $C^{I'}$ ar
the canonical chiral fields and $Q=US$ reduces the Kahler metric $Z_{IJ}$ to the unit
matrix as discussed in Eqs.(18) and (19) above. The soft breaking parameters $A_{IJK}$
are\cite{low4}\begin{equation}\label{eq46}
  A_{IJK}=F^a[\frac{1}{2}\partial_a KY_{IJK}-\Gamma^N_{a(I}Y_{JK)N}+\partial_a
  Y_{IJK}]
\end{equation}
where
\begin{equation}\label{eq47}
  F^a=\kappa_4^2e^{\frac{K}{2}}K^{a\bar b}(\partial_{\bar b}W+W\partial_{\bar b} K),
\end{equation}$K_{a\bar b}=\partial_a\partial_{\bar b}K$ and W is the superpotential.
The first term of Eq. (47) gives rise to the universal soft breaking with
\begin{equation}\label{eq48}
  A_0=\frac{1}{2}F^a\partial_a K
\end{equation}
while the second and third terms give non-universal corrections mixing the generations. The mixing is
significant only between the second and third generations since the S matrix elements of
Eq.(19), which enters into Eq. (46) are large then since $\lambda_{2,3}$ are small. One can
estimate the expected size for $A_{23}^{u,d}$ and finds qualitatively it is about the right
size to account for the B factory data, i.e. Eq.(45)\cite{abh2}. So it possible that this
Horava-Witten model can account for the anomalous B data.

\section{Conclusions}
With the large number of string vacuua and the large number of different string models one
might construct, it is of interest to see if one can find string models that can explain
naturally some aspects of the Standard Model that are a priori arbitrary (and unreasonable).
String theory offers mechanisms to do this that are not available in field theory  GUT
models. We have considered here the Horava-Witten M-theory with the assumption that the
5-branes in the bulk cluster near the distant brane ($d_n=1-z_n\simeq 0.1$) and the
topological parameters $\beta^{(0)}_i$ of the physical brane vanish. This later condition
is non-trivial, but allows a three generation model to exist for a Calabi-Yau
compactification with torus fibration with two sections and a del Pezzo $dP_7$ base. The
parameter $d_n\simeq0.1$ then offers a natural parameter to simultaneously help converge
the Witten $\epsilon$ expansion and account for the quark and lepton mass hierarchy without
significant fine tuning. Thus in this model, the quark and lepton mass hierarchies arise
due to the geometrical structure of the model, i.e. the positions of the 5-branes in the bulk. Similarly, the Kahler potential offers an
alternate way of introducing neutrino masses, different from the field theoretic see-saw
mechanism. The model is distinguishable from the see-saw mechanism in that it gives rise to Dirac
neutrino masses and hence no neutrinoless double beta decay. In addition, it predicts that $U_{e3}$
is small, Eq. (31). One can also see the possible origin of non-universal soft breaking terms.

Models of this type are of course not complete. One can not actually calculate Yukawa
couplings as these involve integrals over a Calabi-Yau manifold which in general can not be
explicitly carried out. (One can only attempt to qualitatively estimate them) While there has been recent work to stabilize string theory
moduli\cite{buchbinder,linde,becker}, one needs an explicit mechanism for stabilizing the 5-branes close to the
distant orbifold  plane which would give a fundamental  
explanation for the phenomenological choice of $d_n\simeq 0.1$. What one can hope models of
this type can do at present give a suggestion as to the direction of where a good
string model might be. 

\end{document}